\documentclass[10pt,a4paper,journal]{IEEEtran}
\IEEEoverridecommandlockouts
\usepackage[a4paper,
            left=0.5in,right=.5in,top=0.6in,bottom=0.75in]{geometry}
\usepackage{graphicx}
\usepackage{color}
\usepackage{cite}
\usepackage{amsmath}
\usepackage[caption=false]{subfig}
\usepackage{mathtools}
\usepackage{amssymb}
\usepackage{amsfonts}

\usepackage{algorithmic}
\usepackage{algorithm}

\usepackage{amsmath}

\usepackage{url}
\newcommand{\subparagraph}{}
\usepackage{amssymb}
\usepackage{amsfonts}
\usepackage{lineno}
\usepackage{comment}

\usepackage{siunitx}

\usepackage{paralist} % Beware of enu­mer­ated and item­ized lists. Instead, replace them with com­pact lists.

\newcounter{tempEquationCounter}
\newcounter{thisEquationNumber}

\begin{document}
%\IEEEpubidadjcol
%\IEEEpubid{1089-7798 (c) 2016 IEEE. Personal use is permitted, but republication/redistribution requires IEEE permission. See http://www.ieee.org/publications_standards/publications/rights/index.html for more information.}
\IEEEpubid{\makebox{1089-7798 (c) 2016 IEEE. Personal use is permitted,  but republication/redistribution requires IEEE permission. \hfill }
\hspace{\columnsep}\makebox{\hfill }}

\vspace{-6.0cm}
\title{Ultra-Reliable and Low Latency Communication in mmWave-Enabled Massive MIMO Networks}\vspace{-2.0cm}
\author{Trung Kien Vu,~\IEEEmembership{Student Member,~IEEE,}
        Chen-Feng Liu,~\IEEEmembership{Student Member,~IEEE,}\\
        Mehdi Bennis,~\IEEEmembership{Senior Member,~IEEE,}
        M\' erouane Debbah,~\IEEEmembership{Fellow,~IEEE,}
        Matti Latva-aho,~\IEEEmembership{Senior Member,~IEEE,}
        and~Choong Seon Hong,~\IEEEmembership{Senior Member,~IEEE}
\vspace{-1cm}
\thanks{Manuscript received March 24, 2017; revised May 05, 2017; accepted May 12, 2017. Date of publication May 17, 2017; This work was supported in part by the Finnish Funding Agency for Technology and Innovation (Tekes),  Nokia,  Huawei,  Anite,  in part by the Academy of Finland CARMA project, in part by the Academy of Finland funding through the grant 284704, and in part by the ERC Starting Grant 305123 MORE (Advanced Mathematical Tools for Complex Network Engineering).}

\thanks{T. K. Vu, C.-F. Liu, M. Bennis, and M. Latva-aho are with the Centre for Wireless Communications, University of Oulu, Oulu 90014, Finland (e-mail: trungkien.vu@oulu.fi; chen-feng.liu@oulu.fi; mehdi.bennis@oulu.fi; matti.latva-aho@oulu.fi).}

\thanks{M. Debbah is with the Large Networks and System Group (LANEAS), CentraleSup\'elec, Universit\'e Paris-Saclay, Gif-sur-Yvette, France and is with the Mathematical and Algorithmic Sciences Laboratory, Huawei France R\&D, Paris, France  (e-mail: merouane.debbah@huawei.com).}

\thanks{C. S. Hong is with the Department of Computer Engineering, Kyung Hee University, Yongin 446-701, South Korea (email: cshong@khu.ac.kr).}

\thanks{Citation information: DOI 10.1109/LCOMM.2017.2705148, IEEE Communications Letters} }\vspace{-6.0cm}
 
\maketitle
\vspace{-10.0cm}
\begin{abstract}
%\boldmath
Ultra-reliability and low-latency are two key components in $5$G networks. In this letter, we investigate the problem of ultra-reliable and low-latency communication (URLLC) in millimeter wave (mmWave)-enabled massive multiple-input multiple-output (MIMO) networks. The problem is cast as a network utility maximization subject to probabilistic latency and reliability constraints. To solve this problem, we resort to the  Lyapunov technique whereby a  utility-delay control approach is proposed, which adapts to channel variations and queue dynamics. Numerical results demonstrate that  our proposed approach ensures reliable communication with a guaranteed probability of $99.99\%$, and reduces latency by $28.41\%$ and $77.11\%$ as compared to \textit{baselines} with and without probabilistic latency constraints, respectively.
\end{abstract}

\IEEEpeerreviewmaketitle
\vspace{-0.2cm}
\section{Introduction} %due to the exponential increase in the number of mobile broadband subscribers single digit ms
\label{Intro}
\IEEEPARstart{C}{urrently}, millimeter wave (mmWave) and massive multiple-input multiple-output (MIMO) techniques are investigated to provide reliable communication with an over-the-air latency of few milliseconds and extreme throughput~\cite{Nokia2011}. While massive MIMO with large degrees of freedom provides high energy and spectral efficiency~\cite{2014massive}, mmWave frequency bands provide large bandwidth \cite{mmwave_GC_16}. In addition, due to the short wavelength of mmWaves, large antenna array can be packed into highly directional beamforming, which makes massive MIMO practically feasible \cite{Vu2016}. Thus far, most of existing works on mmWave-enabled massive MIMO systems focus mainly on providing capacity improvement~\cite{Vu2016}, while latency and reliability are not addressed. Although latency and reliability are applicable to many scenarios (e.g. mission-critical applications), in this work, we are interested in the integration of mmWave communication and massive MIMO techniques, which holds the promise of providing great enhancements of the overall system performance~\cite{Nokia2011, 2014massive, Vu2016}. Specifically, this letter is concerned with addressing the fundamental question in mmWave-enabled massive MIMO systems: ``\emph{how to simultaneously provide order of magnitude capacity improvements and latency reduction}?"
By invoking the Lyapunov optimization framework, an utility-optimal solution is obtained to maximize network throughput subject to queuing stability \cite{neely2010S}. This solution establishes a utility-delay tradeoff, which achieves utility-optimality at the price of large queuing delays \cite{neely2010S}. To cope with this shortcoming, in this letter the Lyapunov framework is extended to incorporate probabilistic latency and reliability constraints, which takes into account queue length, arrival rate, and channel variations with a guaranteed probability. To do so, the problem is formulated as a network utility maximization (NUM). By applying the  drift-plus-penalty technique, the problem is decoupled into a dynamic latency control and rate allocation. Here, the latency control problem is a difference of convex (DC) programming problem, which is solved efficiently by the convex-concave procedure (CCP)~\cite{lipp2014CCP}. Finally, a performance evaluation is carried out to show the latency reduction and the tradeoff between reliability, traffic intensity, and user density.
\vspace{-0.2cm}
\section{System Model}
\label{lb-SM-P}
Consider the downlink (DL) transmission of a single cell massive MIMO system\footnote{Our model can be extended to multi-cell massive MIMO systems in which the problem of inter-cell interference can be addressed by designing a hierarchical precoder at the MBS~\cite{Liu2014}, to mitigate both intra-cell and inter-cell interference, or by applying an interference coordination approach~\cite{2017mmwave}.} consisting of one  macro base station (MBS) equipped with $N$ antennas, and a set, $\mathcal{M} = \{1, \ldots, M\}$, of single-antenna user equipments (UEs). We assume that $N\geq M$ and $N\gg 1$. Further, the co-channel time-division duplexing (TDD)  is considered in which the MBS estimates channels via the uplink phase. We denote  the  propagation channel   between the MBS and the $m{\text{th}}$ UE  as $\mathbf{h}_{m}= \sqrt{N} \mathbf{\Theta}_{m} ^{1/2}\mathbf{\tilde{h}}_{m}$, where $\mathbf{\Theta}_{m} \in \mathbb{C}^{N \times N}$ depicts the antenna spatial correlation, and the elements of $\mathbf{\tilde{h}}_{m}\in \mathbb{C}^{N \times 1}$ are independent and identically distributed (i.i.d.) with zero mean and  variance $1/N$. In addition, channels experience flat and block fading, and imperfect channel state information (CSI) is assumed. As per \cite{wagner2012l}, the estimated channel can be modeled as
$\mathbf{\hat{h}}_{m}=  \sqrt{1 - \tau_m^2} \mathbf{h}_{m}+ \tau _m \sqrt{N} \mathbf{\Theta}_{m} ^{1/2}\mathbf{z}_{m},\forall \,m\in\mathcal{M}.$
Here, $\mathbf{z}_{m}\in\mathbb{C}^{N \times 1}$ denotes the estimated noise vector which has i.i.d.~elements with zero mean and variance $1/N$, and $\tau_m\in[0,1]$ reflects the estimation accuracy; in case of perfect CSI, $\tau _m = 0$.
\IEEEpubidadjcol
Given the estimated channel matrix $\mathbf{\hat{H}} = [ \mathbf{\hat{h}}_{1}, \cdots,
\mathbf{\hat{h}}_{M}]\in \mathbb{C}^{N  \times M}$, the MBS utilizes the regularized zero-forcing\footnote{Other hybrid beamforming designs are left for future works.} (RZF) precoder with the precoding matrix, $\mathbf{V} = [\mathbf{v}_1, \cdots, \mathbf{v}_{M}] \in \mathbb{C}^{N \times M}$, which is given by  $\mathbf{V} = \big( \mathbf{\hat{H}}^{\dag}  \mathbf{\hat{H}} + N \alpha \mathbf{I}_{N}\big)^{-1}  \mathbf{\hat{H}}^{\dag}$. Note that  $\mathbf{v}_{m}$  is the precoding vector for UE $m$. In $\mathbf{V}$, $\dag$ denotes the conjugate transpose, and the regularization parameter $\alpha > 0$ is scaled by $N$ to ensure the matrix $ \mathbf{\hat{H}}^{\dag} \mathbf{\hat{H}} + N \alpha \mathbf{I}_{N}$ is well-conditioned as $N \to \infty$~\cite{Liu2014}. Further, transmit power $p_{m}\geq 0$ is allocated to UE $m$. Denoting  all allocated powers in the diagonal matrix $\mathbf{P} = \text{diag}(p_1, \cdots, p_{M} )$, we have the constraint $\text{Tr}\big (\mathbf{P} \mathbf{V}^{\dag} \mathbf{V} \big) \leq P$, with $P$ the maximum transmit power of the MBS. With the aid of the results in \cite[Theorem 1]{wagner2012l}, the transmit power allocation constraint can be expressed as

\begin{equation}\label{Power constraint}
\textstyle \frac{1}{N}  \sum \limits_{m = 1}^{M} \frac{ {p}_{m}^{}}{  \Omega_{m} } \leq P,\mbox{~and~}p_m\geq 0,~\forall\,m\in\mathcal{M},
 \end{equation}
where the parameter $\Omega_m$ is the solution to $\Omega_{m} = \textstyle \frac{1}{N} \mathrm{Tr}\big (\mathbf{{\Theta}}_m \big( \frac{1}{N} \sum _{m = 1}^{M} \frac{ \mathbf{{\Theta}}_m }{ \alpha + \Omega_{m}} + \mathbf{I}_{N} \big)^{-1}\big)$.
By designing the  precoding matrix $\mathbf{V}$ and transmit power $\mathbf{P}$, the ergodic DL rate of UE $m \in \mathcal{M}$ is
\begin{equation} \label{Rate-UE1}
\textstyle r_{m} (\mathbf{P})  =  \mathbb{E} \Big[  \log_2 \Big( 1 + \frac {   p_{m}^{} |\mathbf{h}_{m}^{ \dag} \mathbf{v}_{m}^{} |^2  } {  1+  \sum _{k = 1, k \neq m}^{M } p_{k}^{} | \mathbf{h}_{m}^{ \dag}
\mathbf{v}_{k}^{} |^2  } \Big)\Big].
\end{equation}
Here, we  invoke results from random matrix theory in order to get the deterministic equivalence for \eqref{Rate-UE1} \cite{wagner2012l}. In particular, as $N\geq M$ and $N\gg 1$, for small $\alpha$,  the ergodic DL rate (\ref{Rate-UE1}) {\emph{almost surely}} converges to
\begin{equation}
\label{SINR-UE-3}
\textstyle r_{m}(\mathbf{P}) \xrightarrow{a.s.}  \log_2 \big( 1 + p_{m} (1 - \tau_m^2)\big), ~\forall \, m \in \mathcal{M},
\end{equation}
where $\xrightarrow{a.s.}$ denotes \emph{almost sure} convergence~\cite{Liu2014}, \cite[Theorem 2]{wagner2012l}. Moreover, we assume that the MBS has queue buffers to store UE data \cite{neely2010S}. In this regard,  we first index the coherence time block by slot $t\in\mathbb{Z}^{+}$. At the  beginning of each slot $t$,  the queue length for UE $m$ is denoted by $Q_m(t)$ which evolves as follows:
% We further note that  the asymptotic expression is sufficiently precise when $N\geq M$ and $N\gg 1$~\cite{wagner2012l}
%
%
\begin{equation}\label{queueQ}
{Q}_{m}(t + 1) = [{Q}_{m}(t) - {r}_{m}(t)]^{+} + {a}_{m}(t), ~ \forall \, m \in \mathcal{M},
\end{equation}
where $[x]^{+}\triangleq\max\{x, 0\}$, and $a_m(t)$ is the data arrival rate of UE $m$. Further, we assume that $a_m(t)$ is i.i.d.~over time slots with mean arrival rate $\lambda_m$ and upper bounded by $a_m^{\max}$ \cite{neely2010S}.
\vspace{-0.2cm}
\section{Problem Formulation}
\label{Op-Form}
According to Little's law \cite{2008little}, the average delay is proportional to $\lim_{T\to\infty}\frac{1}{T}\sum_{t=1}^{T}\mathbb{E}[Q_m(t)]/\lambda_m$. Thus, we use $Q_m(t)/\lambda_m$ as a delay measure and enforce an allowable upper bound $d^{\rm th}_m$.
We further note that the delay  (or queue length) bound violation is related to reliability. Thus, taking into account the latency  and reliability requirements, we characterize the delay bound violation with a tolerable probability. Specifically, we impose a probabilistic constraint on the queue size length for  UE $m \in \mathcal{M}$ as follows:
%~\cite{weiner2014design}
%
%
\begin{equation}\label{delayconst1}
\textstyle\text{Pr}\Big\{ \frac{{Q}_m(t) }{ \lambda_m }  \geq {d}_m^{\text{th}} \Big\} \leq \epsilon_m, ~ \forall \,t.
\end{equation}
In \eqref{delayconst1}, ${d}_m^{\text{th}}$ reflects the UE delay requirement. Here, $\epsilon_m \ll 1$ is the target probability for reliable communication.

In order to reduce latency, the intuitive way is to send as many data as possible. However, this might over-allocate resources to UEs, i.e., $r_m(t)\gg Q_m(t)$. To handle this issue, we enforce a maximum rate constraint $r_m^{\mathrm{max}}$ for each UE $m$. On the other hand, we enforce the MBS to guarantee for all UEs a certain level of QoS, i.e., the minimum rate requirement $r_m^{\mathrm{min}},\forall\,m\in\mathcal{M}$.

We define the  network utility as  $\sum _{m = 1}^{M} \omega_{m}f( \bar{r}_{m} )$ where  the time average expected rate
$\bar{r}_{m}=\lim_{T \to \infty} \frac{1}{T}  \sum_{t=1}^{T} \mathbb{E} [ r_{m}(t) ]$ and the non-negative weight $\omega_{m}$ for each UE $m$.
Additionally, we assume that $f(\cdot)$ is a strictly concave, increasing, and twice continuously-differentiable function. Taking into account these constraints presented above yields the following network utility maximization:
\begin{subequations}\label{eq:Obj-Formulate-0}
\begin{IEEEeqnarray}{rcl}
 \hspace{-2em}  {\mbox{\bf{OP}}}:~&\underset{ \mathbf{P}(t) }{\text{max}}
        & \quad \textstyle \sum\limits _{m = 1}^{M} \omega_{m} f( \bar{r}_{m} )\label{eq-01}
  \\&\text{subject to}
  &\quad r_m^{\mathrm{min}} \leq {r}_{m}(t) \leq r_m^{\mathrm{max}}, ~ \forall \, m \in \mathcal{M}, \,\forall\,t,\label{admisconst}
\\        && \quad\eqref{Power constraint}\mbox{~and~}\eqref{delayconst1}.\notag
\end{IEEEeqnarray}
\end{subequations}
Our main problem involves a probabilistic constraint \eqref{delayconst1}, which cannot be addressed tractably. To overcome this challenge, we apply Markov's inequality \cite{linearize_Q_2013} to linearize \eqref{delayconst1} such that
%
%
%
%\begin{equation}
%\label{delayconst2}
$\text{Pr}\big\{ \frac{{Q}_m(t) }{ \lambda_m }  \geq {d}_m^{\text{th}} \big\}\leq \frac{ \mathbb{E}[Q_m(t) ]}{{\lambda_m {d}_m^{\text{th}}  }}$.
%\end{equation}
%
%
%
Then, \eqref{delayconst1} is satisfied if
\begin{equation}\label{delayconst3}
 \textstyle\mathbb{E} [{Q}_m(t)   ] \leq \lambda_m {d}_m^{\text{th}} \epsilon_m, ~ \forall \,m \in \mathcal{M},\,\forall\,t.
\end{equation}
 Thereafter, we consider  \eqref{delayconst3} to represent the latency and reliability constraint. Assuming that $\{{a}_m(t)|\forall\,t\geq 1\}$ is a Poisson arrival process \cite{linearize_Q_2013}, we have that $\mathbb{E} [ {Q}_m(t) ] = t\lambda_m -\sum_{\tau=1}^{t}r_m(\tau)$ which is plugged into \eqref{delayconst3}. Subsequently, we obtain
\begin{multline}
\hspace{-0.5em} \textstyle  r_m(t) \geq  t\lambda_m - \lambda_m {d}_m^{\text{th}} \epsilon_m-\sum\limits_{\tau=1}^{t-1}r_m(\tau),~\forall\, m \in \mathcal{M},\,\forall\,t,\label{delayconst5}
\end{multline}
which represents the minimum rate requirement in slot $t$ for UE $m$ for reliable communication.
Here, we transform the probabilistic latency and reliability constraint \eqref{delayconst1} into one linear constraint \eqref{delayconst5} of instantaneous rate requirements, which helps to analyse and optimize the URLLC problem.  In particular, if the delay requirement/reliability constraint is looser (i.e., larger $d^{\rm th}_m$ or $\epsilon_m$),  the instantaneous rate requirement is reduced. In contrast, if we have a tighter constraint for reliable communication or delay requirement, then the instantaneous rate requirement is higher. Combining \eqref{admisconst} and  \eqref{delayconst5}, we rewrite {\bf OP} as follows:
\begin{subequations}\label{eq:Obj-Formulate-1}
%\label{eq:Obj-Formulate-0} %- \alpha\, {c}_{0} ( \bar{\mathbf{Q}},  \bar{\mathbf{a}} )
\begin{IEEEeqnarray}{cl}
   \hspace{-2em} \underset{ \mathbf{P}(t) }{\text{max}}
        & \quad\textstyle   \sum\limits_{m = 1}^{M} \omega_{m} f( \bar{r}_{m} ) %\label{eq-01}\\
  \\ \hspace{-2em}\text{subject to}
      &\quad r_{m}^{0}(t) \leq r_m(t) \leq{r}_{{m}}^{\mathrm{max}},~\forall\, m \in \mathcal{M},\,\forall\,t,
      \\  & \quad  \mbox{and~}\eqref{Power constraint} ,\notag
        % \label{eq-03}
\end{IEEEeqnarray}
\end{subequations}
with $r_{m}^{0}(t) = \max \{{r}_{{m}}^{\mathrm{min}},  t\lambda_m - \lambda_m {d}_m^{\text{th}} \epsilon_m-\sum_{\tau=1}^{t-1}r_m(\tau)\}$.

\vspace{-0.2cm}
\section{ Lyapunov Optimization Framework}
\label{LOF}
To tackle  \eqref{eq:Obj-Formulate-1}, we resort to Lyapunov optimization techniques \cite{neely2010S}.
Firstly, for each DL rate $r_m(t)$, we introduce the auxiliary variable vector $\boldsymbol{\varphi}(t)=(\varphi_{m}(t)|\forall\,m\in\mathcal{M})$ that satisfies
\begin{align}
&\textstyle\bar{\varphi}_{m}=  \lim\limits_{T \to \infty} \frac{1}{T}   \sum\limits_{t=0}^{T} \mathbb{E} \big[ \varphi_{m}(t)  \big]  \leq \bar{r}_{m},~\forall\,m\in\mathcal{M},\label{average_auxiliary_constraint}
\\&\varphi_{m}^{0}(t) \leq \varphi_{m}(t) \leq {r}_{m}^{\mathrm{max}},~\forall \, m \in \mathcal{M}, \,\forall\, t ,\label{auxiliary_bound}
\end{align}
with $\varphi_{m}^{0}(t) = \max \{{r}_{{m}}^{\mathrm{min}},t\lambda_m - \lambda_m {d}_m^{\text{th}} \epsilon_m-\sum_{\tau=1}^{t-1}\varphi_m(\tau)\}$.
Incorporating the auxiliary variables, \eqref{eq:Obj-Formulate-1} is equivalent to
\begin{IEEEeqnarray*}{rcl}
{\mbox{\bf LP}}:\quad  & \underset{ \mathbf{P}(t),\boldsymbol{\varphi} (t)}{\text{max}}
        & \quad   \textstyle\lim\limits_{T\to\infty}\frac{1}{T} \sum \limits_{t=1}^{T}\sum\limits _{m = 1}^{M} \omega_{m} \mathbb{E}[f(\varphi_m(t))] \label{eq-1a} \\
   &\text{subject to}
&\quad  \eqref{Power constraint} ,\,\eqref{average_auxiliary_constraint},\,\mbox{and }\eqref{auxiliary_bound}.\notag
        %& \quad  \varphi_m(t) \geq  a_m ( t - \beta_m ) - \sum_{\tau = 0}^{t-1} \varphi_m(\tau), \forall m \in \mathcal{M},\label{eq-1cc} \\
\end{IEEEeqnarray*}
%\end{subequations}
%
%
%
In order to ensure the inequality constraint \eqref{average_auxiliary_constraint}, a virtual queue vector $\mathbf{Y(t)} = (Y_m(t)|\forall\,m \in \mathcal{M})$ is introduced, where each element evolves according to
\begin{equation}\label{queueY}
{Y}_{m}(t + 1) = [{Y}_{m}(t) + \varphi_{m}(t) - {r}_{m}(t)]^{+},~ \forall \, m \in \mathcal{M}.
\end{equation}
Subsequently, we express the conditional Lyapunov drift-plus-penalty for each time slot $t$ as:
\begin{multline}\label{drift_plus_penalty}
  \textstyle\mathbb{E} \bigg [ \sum\limits _{m = 1}^{M}  \Big[ \frac{1}{2} {Y}_{m}(t+1)^{2}   - \frac{1}{2} {Y}_{m}(t)^{2}
  \\-\nu_m(t) w_m  f( {\varphi}_{m}(t) )\big]  \big|\mathbf{Y}(t)\Big].
\end{multline}
In \eqref{drift_plus_penalty}, $\nu_m(t)$ is the control parameter which affects the utility-queue length tradeoff. This control parameter is conventionally chosen to be static and identical for all UEs \cite{neely2010S}. However, this setting does not hold for system dynamics (e.g., instantaneous data arrivals) and the diverse system configuration (i.e., different delay and QoS requirements). Thus, we dynamically design these control parameters.
From the analysis in the Lyapunov optimization framework~\cite{neely2010S}, we can find  ${Y}_{m}(t)  \leq  \nu_m(t) \omega_m \pi_m + a_{m}^{\mathrm{max}}$ with $\pi_m$ being the largest first-order derivative of $f( x )$. Letting $\omega_m= 1, \forall\, m \in \mathcal{M}$, we have the lower bound
%
%
%
%\begin{equation}\label{Bound6}
$\pi_m\nu_m(t)     \geq  \nu_m^{0}(t), \forall\, m \in \mathcal{M}$,
%\end{equation}
%
%
%
for selecting the control parameters, where $ \nu_m^{0}(t) = \max \{  {Y}_{m}(t) - a_{m}^{\mathrm{max}} ,  1\}$.
Subsequently, following the straightforward calculations of the Lyapunov drift-plus-penalty technique which are omitted  for space, we obtain
\begin{subequations}
\begin{align}
\eqref{drift_plus_penalty}\leq   \textstyle \mathbb{E} \bigg [ \sum\limits _{m = 1}^{M} \big({Y}_{m}(t) {\varphi}_{m}(t) -  \nu_m(t) \omega_{m} f \big( {\varphi}_{m}(t) \big) \big)&\label{Lyapunob_upperbound_1}
\\ \textstyle-  \sum\limits_{m = 1}^{M} {Y}_{m}(t) {r}_{m} \big(\mathbf{P}(t)   \big)+C  \big|\mathbf{Y}(t)\Big]&.\label{Lyapunob_upperbound_2}
\end{align}
\end{subequations}
Due to  space limitation, we omit the details of the constant value $C$ which does not influence the system performance~\cite{neely2010S}.
We note that the solution to {\bf LP} is acquired by minimizing the right-hand side (RHS) of \eqref {Lyapunob_upperbound_1} and \eqref{Lyapunob_upperbound_2} in every slot $t$. Further, \eqref {Lyapunob_upperbound_1} is related to the reliability and QoS requirements while \eqref{Lyapunob_upperbound_2} reflects optimal power allocation to UEs.
\vspace{-0.3cm}
\subsection{Auxiliary Variable and Control Parameter  Selection}
\label{AV}
Considering the logarithmic fairness utility function, i.e., $f(x)=\log (x)$, minimizing the RHS of \eqref{Lyapunob_upperbound_1} for each $m\in\mathcal{M}$ is formulated as:
\begin{subequations}\label{eq:Optimal-AR}
\begin{IEEEeqnarray}{cl}
   \hspace{-2em} \underset{ \varphi_m(t),\,\nu_m(t) }{\text{min}}
        &  \quad  {Y}_{m}(t) \varphi_{m}(t)-  \nu_m(t)  \log \big( {\varphi}_{m}(t)\big)\label{eq-obj2}\\
\hspace{-2em}  \text{subject to}
    & \quad \pi_m \nu_m(t)   \geq  \nu_m^{0}(t),\label{eq-c4} \\
    & \quad r_{m}^{0}(t) \leq \varphi_{m}(t) \leq {r}_{m}^{\mathrm{max}}.\label{eq-1c}
\end{IEEEeqnarray}
\end{subequations}
Before proceeding with problem \eqref{eq:Optimal-AR}, we rewrite  $-  \nu_m(t)  \log ( {\varphi}_{m}(t))$
% the second term of
in \eqref{eq-obj2}, for any $\varphi_m(t) > 0$ and $\nu_m (t)> 0$, as
\begin{equation}
  % -  \nu_m(t)  \log \bigg( \frac{{\varphi}_{m}(t)}{\nu_m(t)} \nu_m(t)\bigg)  \\=
   \underbrace{      \nu_m(t) \log \bigg( \frac{\nu_m(t) }{{\varphi}_{m}(t)} \bigg)}_{ h_{0}({\varphi_m,\,\nu_m})}  -\underbrace{\nu_m(t)  \log \big( \nu_m(t)\big)}_{g_{0}({\nu_m})}, \notag %\label{eq-DC}
\end{equation}
in which both $h_{0}({\varphi_m,\nu_m})$ (i.e., relative entropy function) and $g_{0}({\nu_m})$ (i.e., negative entropy function) are convex functions.
Since \eqref{eq-obj2} is the difference of convex functions while constraints \eqref{eq-c4}  and \eqref{eq-1c} are affine functions, problem \eqref{eq:Optimal-AR} belongs to DC programming problems~\cite{Le2005}, which can be efficiently and iteratively addressed by the CCP~\cite{lipp2014CCP}.
The CCP algorithm to obtain the solution to problem \eqref{eq:Optimal-AR} is detailed in Algorithm~\ref{algUS}.
We note that the CCP  provably converges to the local optima of DC programming problems \cite{lipp2014CCP}. However, due to space limitation, we omit the convergence proof of Algorithm~\ref{algUS}  (please refer to~\cite{lipp2014CCP} for the formal proof).
% We will briefly describe the convergence here for the sake of completeness since it was studied in~\cite{lipp2014CCP}.
% Moreover, we omit the convergence here for the sake of completeness since it was studied in~\cite{lipp2014CCP}.}
%
%
\begin{algorithm}[!t]% enter the algorithm environment
\caption{CCP algorithm for solving sub-problem \eqref{eq:Optimal-AR}.}% give the algorithm a caption% and a label for \ref{} commands later in the document
\label{algUS}
\begin{algorithmic}
    \WHILE {$m \in \mathcal{M}$}
    \STATE Initialize $i = 0$ and a feasible point $\nu_{m}^{(i)}$ in \eqref{eq-c4}.
    \REPEAT
        \STATE $\text{Convexify}$ ~$\hat{g}_{0} (\nu_m, \nu_{m}^{(i)}) = g_{0}(\nu_{m}^{(i)}) + \nabla g_{0}(\nu_m -\nu_{m}^{(i)})$.
        \STATE $\text{Solve:}$  \\
              $~~~~\underset{\varphi_m,\nu_m}{\mbox{min}}~~ ~~h_{0}(\varphi_m,\nu_m) - \hat{g}_{0}(\nu_m, \nu_{m}^{(i)}) + {Y}_{m} \varphi_{m} $  \\
              $~~\text{subject to} ~\eqref{eq-c4}\mbox{ and }\eqref{eq-1c},$ \\
              Find the optimal $\varphi_m^{(i)\star}$ and $\nu_m^{(i)\star}$. \\
        \STATE $\text{Update}$ $\nu_m^{(i+1)}\coloneqq \nu_m^{(i)\star}$ and $i \coloneqq i + 1$.
    \UNTIL{\text{Convergence}}
    \ENDWHILE
\end{algorithmic}
\end{algorithm}
\vspace{-0.3cm}
\subsection{Power Allocation}
\label{PowerControl}
The optimal transmit power in \eqref{Lyapunob_upperbound_2} is computed by
\begin{subequations}
\begin{IEEEeqnarray}{cl}
    \underset{{\mathbf{P}}(t)}{\text{min}}
        & \quad  \textstyle -\sum\limits_{m = 1}^{M} {Y}_{m}(t) r_{m}(\mathbf{P}(t))\notag
   \\ \text{subject to}
        & \quad \eqref{Power constraint}.\notag
\end{IEEEeqnarray}
\end{subequations}
Here, the objective function is strictly convex for ${p}_{m}(t) \geq 0, \forall \,m \in \mathcal{M}$, and the constraints are compact. Therefore, the optimal solution of $\mathbf{P}^{\star}(t)$ exists and is efficiently reached by numerical methods.

After obtaining the optimal auxiliary variable and transmit power, we  update the queues ${Q}_m(t+1)$  and  ${Y}_m(t+1)$  as per \eqref{queueQ} and \eqref{queueY}, respectively.
\vspace{-0.3cm}
\section{Numerical Results} %
\label{Evaluation}
We consider a single-cell massive MIMO system\footnote{The multi-cell scenario raises a problem of additional delay due to the need of information exchange among base stations, which is required by either the coordination scheme or distributed approach. This problem is also a very interesting open topic for future work.} in which the MBS, with $N= 32$ antennas and $P= 38$\,dBm, is  located at the center of the $0.5 \times 0.5$ $\text{km}^{2}$ square area. UEs (from $8$ to $60$ UEs per $\text{km}^2$) are randomly deployed within the MBS's coverage with  a minimum MBS-UE distance of $35$\,m. Data arrivals follow a Poisson distribution with different means, and the rate requirements are specified as $r_{m}^{\max} = 1.2 \lambda_m,r_{m}^{\min} = 0.8 \lambda_m,\forall\,m\in\mathcal{M}$. The system bandwidth is 1\,GHz. The path loss is modeled as a distance-based path loss with the line-of-sight (LOS) model\footnote{We assume that the probability of LOS communication is very high, while the impact of other channel models is left for future works.} for urban environments at $28$\,GHz~\cite{mW2014}.  The maximum delay requirement ${d}^{\text{th}}$ and the target reliability probability $\epsilon$ are set to $ 10\,\text{ms}$ and $5\%$, respectively. The numerical results are obtained via Monte-Carlo simulations over $10 000$ realizations with different channel realizations and UE locations.  Furthermore, we compare our proposed scheme with the following baselines:
\begin{itemize}
  \item \textit{Baseline} $1$ refers to  the Lyapunov framework in which the probabilistic latency constraint $\eqref{delayconst1}$ is considered.

  \item \textit{Baseline} $2$ is a variant of \textit{Baseline} $1$ without the probabilistic latency constraint $\eqref{delayconst1}$.
\end{itemize}
\vspace{-0.3cm}
\subsection{Impact of Arrival Rate} %
\label{sub1}
\begin{figure}[t]
	%\hspace{-1.5em}
	\centering
	\includegraphics[width=1\columnwidth]{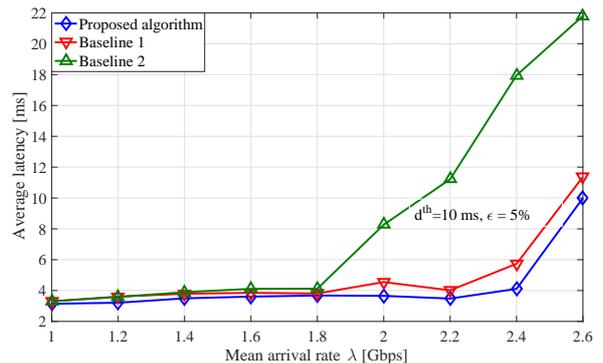}
	\vspace{-2em}
	\caption{Average latency versus mean arrival rates, $M=16$ per $\text{km}^2$.} \label{avgDelay}
    \vspace{-1em}
\end{figure}
In Fig.~\ref{avgDelay}, we report the average latency versus the mean arrival rates $\lambda = \mathbb{E}[a(t)]$ for $M=16$. At low $\lambda$, all schemes do not violate latency constraints, and our proposed algorithm outperforms other \textit{baselines} with a small gap. At higher $\lambda$, the average delay of \textit{baseline} $2$  increases dramatically as $\lambda>1.8$\,Gbps, since \textit{baseline} $2$  does not incorporate the delay constraint, whereas our proposed scheme reduces latency by $28.41\%$ and $77.11\%$ as compared to \textit{baselines} $1$  and $2$, respectively, when $\lambda = 2.4$\,Gbps. When $\lambda > 2.4$\,Gbps, the average delay of all schemes increases, violating the delay requirement of $10$\,ms. It can be observed that under limited maximum transmit power, at very high traffic demand, the latency requirement could not be guaranteed. This highlights the tradeoff between the mean arrival rate and latency.
\begin{figure}[t]
	%\hspace{-1.5em}
\centering
	\includegraphics[width=1\columnwidth]{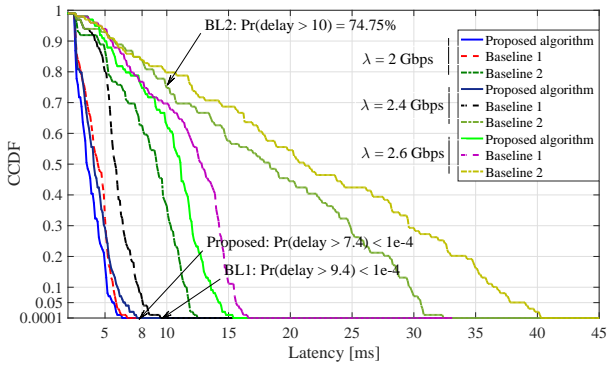} %DelayCCDF_20170322
\vspace{-2em}
	\caption{Tail distribution (CCDF) of latency.}\label{ccdf1}
\vspace{-1em}
\end{figure}
In Fig.~\ref{ccdf1}, we report the tail distribution (complementary cumulative distribution function (CCDF)) of latency to showcase how often the system achieves a delay greater than target delay levels. In particular, at $\lambda = 2.4$\,Gbps, by imposing the probabilistic latency constraint $\eqref{delayconst1}$,  our proposed approach and \textit{baseline} $1$  ensure reliable communication with better guaranteed probabilities, i.e, $\text{Pr}(\text{delay} > 7.5 \text{ms}) < 10^{-4}$ and $\text{Pr}(\text{delay} > 9.4 \text{ms}) < 10^{-4}$, respectively. In contrast, \textit{baseline} $2$  violates the latency constraint with a high probability, where $\text{Pr}(\text{delay} > 10 \text{ms}) = 74.75\%$.
\vspace{-0.5cm}
\subsection{Impact of User Density} %
\begin{figure}[t]
%\hspace{-1.5em}
	 \includegraphics[width=1\columnwidth]{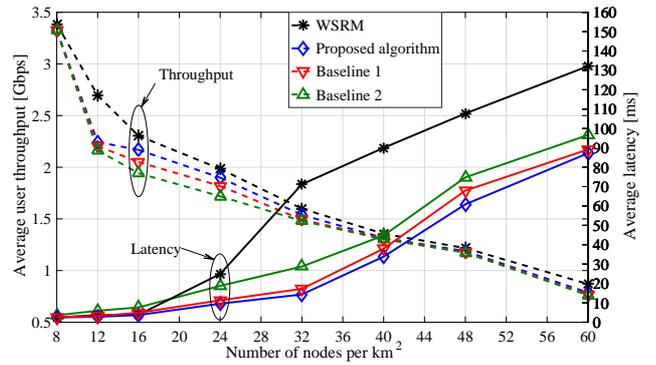}
\vspace{-2em}
	\caption{Average latency and avgUT versus number of users per $\text{km}^2$, $\lambda = 2$ Gbps.}\label{nodedensity}
\vspace{-1em}
\end{figure}

\label{sub2}
In Fig.~\ref{nodedensity}, we compare the average user throughput (avgUT) and average latency of our proposed approach with the two {\it baselines} under the impact of user density. Additionally, we  consider the weighted sum rate maximization (WSRM) case without considering queue  dynamics, i.e., problem \eqref{eq:Obj-Formulate-0} without  the constraints \eqref{delayconst1} and \eqref{admisconst}. The WSRM case is the conventional way to find the system throughput limit but suffers from higher latency. Since all users share the same resources,  the average delay (``solid lines'') increases with the number of users $\emph{M}$, whereas the avgUT (``dash lines") decreases. Fig.~\ref{nodedensity} further shows that when  $\emph{M} > 24$, the delay of all schemes increases dramatically and is far-above the latency requirement. Hence, only a limited number of users can be served  to guarantee the delay requirement, above which, a tradeoff between latency and network density exists.
Our proposed approach achieves better throughput and higher latency reduction than {\it baselines} 1 and 2, while the WSRM case has the worst delay performance as expected.
Compared with WSRM, our proposed approach maintains at least $87\%$  of the throughput limit,  while achieving up to $80\%$ latency reduction. Moreover, our proposed approach reaches Gbps capacity, which represents the capacity improvement brought by the combination of mmWave and massive MIMO techniques. Numerical results show that our approach \emph{simultaneously provides order of magnitude capacity improvements and latency reduction}.
\vspace{-0.2cm}
\section{Conclusion}
\label{Conclusion}
In this letter, we have investigated the problem of mmWave-enabled massive MIMO networks from a latency and reliability standpoint. Specifically, the problem is modeled as a NUM problem subject to the probabilistic latency/reliability constraint and QoS/rate requirement. By incorporating these constraints, we have proposed a dynamic Lyapunov control approach, which adapts to channel variations and system dynamics. Numerical results show that our proposed approach reduces the latency by $28.41\%$ and $77.11\%$ as compared to current \textit{baselines}.
%\vspace{-0.3cm}
\bibliographystyle{IEEEtran}
\bibliography{MassiveMIMO}
\end{document}